\documentstyle[aps,prl,epsfig,multicol]{revtex}
\begin{document}
\tightenlines

\title{Crossover between Thermally Assisted and Pure Quantum Tunneling
in Molecular Magnet Mn$_{12}$-Acetate }

\author{Louisa Bokacheva and Andrew D. Kent}
\address{
Department of Physics, New York University,
4 Washington Place, New York, New York 10003}

\author{Marc A. Walters}
\address{
Department of Chemistry, New York University,
31 Washington Place, New York, New York 10003}

\date{June 19, 2000}

\maketitle

\begin{abstract}
The crossover between thermally assisted and pure quantum tunneling has been
studied in single crystals of high spin ($S=10$) uniaxial molecular magnet 
Mn$_{12}$ using micro-Hall-effect magnetometry. Magnetic hysteresis and 
relaxation experiments have been used to investigate the energy levels 
that determine the magnetization reversal as a function of magnetic field 
and temperature. These experiments demonstrate that the crossover occurs 
in a narrow ($\sim 0.1$~K) or broad ($\sim 1$~K) temperature interval 
depending on the magnitude of the field transverse to the anisotropy axis.
\end{abstract}
\begin{multicols}{2}

High spin molecular magnets Mn$_{12}$ and Fe$_8$ have been actively studied as 
model systems for the behavior of the mesoscopic spins 
\cite{Chudnovsky,Friedman,Ohm,Sessoli,Novak,Hernandez,Wernsdorfer,EPR1,EPR2,EPR3,NS,CCNS}. 
These 
materials can be considered as monodisperse ensembles of weakly interacting
nanomagnets with net spin $S=10$ and strong uniaxial anisotropy. They provide
a unique opportunity to study the interplay between classical thermal 
activation
and quantum tunneling of the magnetization. Of particular interest was 
the observation of a regular series of steps and plateaus in magnetic 
hysteresis loops of Mn$_{12}$ and Fe$_8$ at well defined field intervals 
\cite{Friedman,Ohm}. 
The steps correspond to enhanced relaxation of magnetization, and their 
temperature dependence suggests that both thermal activation and quantum 
tunneling are important to the magnetization reversal \cite{Novak}. Other 
important 
results include the observation of non-exponential relaxation of 
magnetization \cite{Ohm} and quantum phase interference in Fe$_8$ 
\cite{Wernsdorfer}. 
Further, EPR and inelastic neutron scattering experiments have provided 
important information about the magnetic energy levels of Mn$_{12}$ and Fe$_8$ 
and allowed determination of the parameters in an effective spin Hamiltonian 
of these clusters, relevant to understanding their macroscopic magnetic 
response \cite{EPR1,EPR2,EPR3,NS,CCNS}.

Recent theoretical models of spin tunneling suggest that different types of 
crossovers between thermal activation over the anisotropy barrier and quantum 
tunneling under the barrier are possible in the large spin limit 
\cite{Crossover,Crosstheory}. 
The crossover can occur in a narrow temperature interval with the energy at 
which the system crosses the anisotropy barrier shifting abruptly with 
temperature (denoted a first-order crossover), or the crossover can occur 
in a broad interval of temperature with this energy changing smoothly 
with temperature (second-order) \cite{Note}. The ``phase diagram'' for this 
crossover 
depends on the form of the spin Hamiltonian, particularly the terms important 
for tunneling. In finite spin systems the crossover is always smeared. 
Nevertheless, these scenarios are fundamentally different and can be 
distinguished experimentally. In the first case, there are competing maxima 
in the relaxation rate versus energy and the global maximum shifts abruptly 
from one energy to the other as a function of temperature. In the second-order 
case there is a single maximum in the relaxation rate, which shifts 
continuously 
with temperature. Recent experiments have shown that the crossover occurs in 
narrow temperature interval in Mn$_{12}$ when the applied field is parallel 
to the easy axis of the sample \cite{OurEPL}. In contrast, experiments on Fe$_8$ 
suggest a second-order crossover \cite{WernsdorferEPL}. 

In this Letter we show that in Mn$_{12}$ the crossover indeed is 
one in which there are competing maxima in the relaxation rate. We show that 
a transverse 
magnetic field makes the crossover more gradual and leads to a continuous 
shift in the dominant energy levels with temperature (i.e., a second-order 
crossover). Importantly, measurements of the magnetization relaxation as a function of 
temperature also show evidence for a temperature independent regime below 
the crossover temperature. 

Experimental results have been interpreted within an effective spin 
Hamiltonian for an individual cluster:
\begin{equation}
{\cal H}=-DS_z^2 - BS_z^4 - g_z\mu_{\rm B} S_zH_z +{\cal H}',
\label{Ham}
\end{equation}
where the uniaxial anisotropy parameters $D$ and $B$ have been determined by 
EPR \cite{EPR3} and inelastic neutron spectroscopy experiments \cite{NS} 
[$D=0.548(3)$~K, 
$B=1.17(2)\times 10^{-3}$~K, and $g_z$ is estimated to be $1.94(1)$]. Here 
$\cal H'$ includes terms which do not commute with $S_z$ and produce tunneling. 
These mechanisms of level mixing may be due to a transverse field (such as 
hyperfine fields, dipolar fields, or an external field, contributing terms such as 
$H_xS_x$) 
or higher order transverse anisotropies, for example, $C(S_+^4+S_-^4)$, 
$C=2.2(4)\times10^{-5}$~K 
\cite{NS}, which is the lowest-order term allowed by the tetragonal symmetry of 
the Mn$_{12}$ crystal. 
The steps in the hysteresis curves are ascribed to thermally assisted 
tunneling (TAT) or pure quantum tunneling (QT). 
According to this model, the magnetization relaxation occurs 
by tunneling 
from magnetic sublevels ($m=10,9, 8,... ,  -8, -9, -10$), when two 
levels on the opposite sides of the barrier are brought close to resonance by 
the magnetic field. From the unperturbed Hamiltonian (1) the longitudinal ($z$-axis) 
field at which the levels $m_{\rm esc}$ and $m'$ become degenerate is:
\begin{equation}
H(n, m_{\rm esc})=nH_0 \{1+B/D [ m_{\rm esc}^2+(m_{\rm esc}-n)^2 ]\}
\label{StepPos}
\end{equation}
where $n=m_{\rm esc}+m'$ is the step index describing the bias field and 
$H_0=D/g_z\mu_{\rm B}$ 
is a constant ($0.42$ T). The transverse anisotropy does not  
significantly change the resonance fields, as we have checked 
by direct numerical diagonalization of the Hamiltonian (1).

Note that larger magnetic field is necessary to 
bring lower lying sublevels into resonance. As the temperature decreases, the 
thermal population of the excited levels is reduced, and these states 
contribute less and less to the tunneling. Consequently, the steps in hysteresis 
curves shift 
to higher bias field values, and steps with larger $n$ become observable. At low 
temperature, tunneling from the lowest level in the metastable well dominates, and the position 
and amplitude of the steps become independent of temperature, denoted the pure 
quantum tunneling regime (QT).
\vskip0.8cm
\begin{figure}
\begin{center}
\includegraphics[width=8cm,height=8cm]{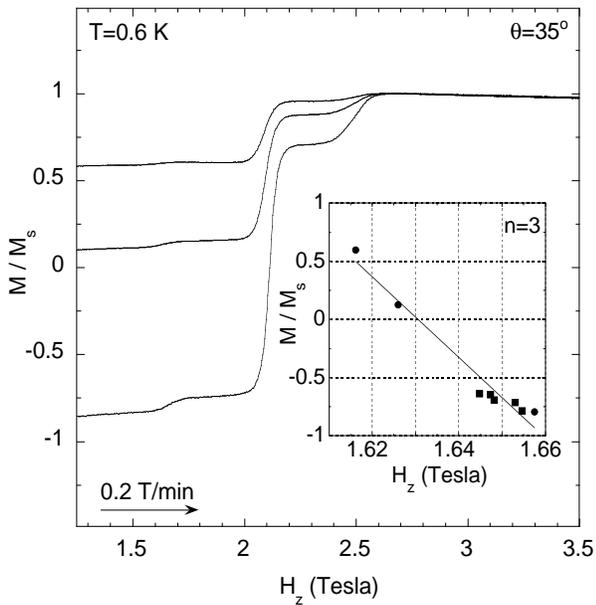}
\vskip0.8cm
\caption{Hysteresis curves of a Mn$_{12}$ single crystal 
measured at $\theta=35^\circ$ 
for three different initial magnetization states: $M_0=0, 0.54 M_s, -M_s$. 
Inset shows the change of the $n=3$ peak position vs magnetization at the step. 
Circles show data points from hysteresis measurements, squares are from field 
sweeps across the peak.}
\label{fig1}
\end{center}
\end{figure}

Our experiments have been conducted using a micro-Hall-effect 
magnetometer \cite{HallBar} in a high field helium 3 system. Single crystals 
of Mn$_{12}$ 
in the shape of parallelepipeds $50 \times 50 \times 200$ $\mu$m$^3$ were 
synthesized according to the procedure described in Ref. \cite{Lis}.  The crystal was 
encapsulated in thermally conducting grease and the temperature was measured with 
a calibrated carbon thermometer a few millimeters from the sample. The angle $\theta$ 
between the easy axis of the crystal and the applied magnetic field was varied 
by rotating the sample in a superconducting solenoid. Three different orientations 
have been studied: $\theta=0^\circ$, $20^\circ$, and $35^\circ$, within an accuracy of 
a few degrees.

Hysteresis curves obtained for $\theta=35^\circ$ are shown on Fig. 1. The sample 
was prepared in three different initial magnetization states: $M_0=0, 0.54M_s, -M_s$, 
by field cooling, then the field was ramped at a constant rate ($0.2$ T/min) towards 
positive saturation. The curves show steps and plateaus, separated by a field 
interval of approximately $0.44$ T, in agreement with previously published 
results. The inset of Fig. 1 shows the field position of the $n=3$ step versus sample 
magnetization at this step. The displayed data were obtained from hysteresis 
measurements such as those shown in Fig. 1 and from measurements in which the 
field was swept back and forth across the step, with the sample magnetization 
varying on each crossing.  The peak positions are seen to depend slightly on 
the sample magnetization due to the average internal dipolar fields. 
Assuming that the peak positions are a linear function of magnetization, 
$H_z = B_z - 4\pi \alpha M_z$, an average $\alpha$, determined from different peaks, 
is approximately $0.51$.

A series of isothermal hysteresis measurements have been performed in small 
intervals of temperature, starting with the sample initially saturated ($M=-M_s$). 
Figure 2 shows a plot of the derivative of magnetization $dM/dH$ versus the 
longitudinal applied field at different temperatures for two orientations, $20^\circ$ 
and $35^\circ$. The positions and structure of the peaks in $dM/dH$ show the magnetic 
fields at which there are maxima in the magnetization relaxation rate at a given 
temperature, applied field, and magnetization. The dashed lines mark the positions 
of the experimental maxima showing their shift with temperature.
Consider the data for $20^\circ$, shown in Fig. 2(a). As the temperature decreases
from $1.34$ to $1.2$~K, the maximum in $dM/dH$ (at $H=1.97$~T) shifts to higher field 
values. At $T=1.24$~K, two high-field shoulders appear, which can be interpreted as the 
``turning on'' of relaxation from energy levels closer to the bottom of the potential well. 
Between $1.34$ and $1.17$~K, amplitude in the lower field 
peaks is reduced, and at $T=1.17$~K the three peaks are of approximately equal 
height. However, when the temperature decreases by $0.03$~K, the maximum shifts 
to the peak which occurs at $H=2.16$~T. 
On lowering the temperature from $1.14$ 
to $0.94$~K, the amplitude of the low-field peaks decreases, which means that 
the tunneling from excited levels is ``frozen out''. 
At $T<1$~K only one maximum at 
$H=2.16$~T survives, and its amplitude and position remain independent of 
temperature down to $0.6$~K, which we associate with pure QT.
 
We can compare the positions of the peaks in this picture with the values of the 
resonant field, calculated according to Eq. (2). The high temperature regime 
corresponds to tunneling mostly from $m_{\rm esc}=8$, for which $H(4,8)=1.97$~T. 
The peaks appearing at higher fields are due to tunneling from $m_{\rm esc}=9$ 
[$H(4,9)=2.06$~T] and $m_{\rm esc}=10$ [$H(4,10)=2.17$~T]. In the pure quantum regime 
the ground state, $m_{\rm esc}=10$, dominates the tunneling. 
The crossover from $m_{\rm esc}=8$ (TAT) to $m_{\rm esc}=10$ (QT) 
occurs over an interval of less than $0.05$~K. 

In contrast with this abrupt crossover, for $\theta=35^\circ$ the peak with the same index 
$n=4$ shifts gradually to the higher field in the range of $1.35-0.75$~K, 
as shown on Fig. 2(b). Below approximately $0.75$~K, the peak remains at a constant 
field value of $2.11$~T, which indicates the transition to the quantum regime. 
In this case the three escape levels, $m_{esc}=8$, $9$, and $10$ are active over 
comparable temperature intervals, which are marked by small steps on the dashed line.

\begin{figure}
\begin{center}
\includegraphics[width=8cm,height=9cm]{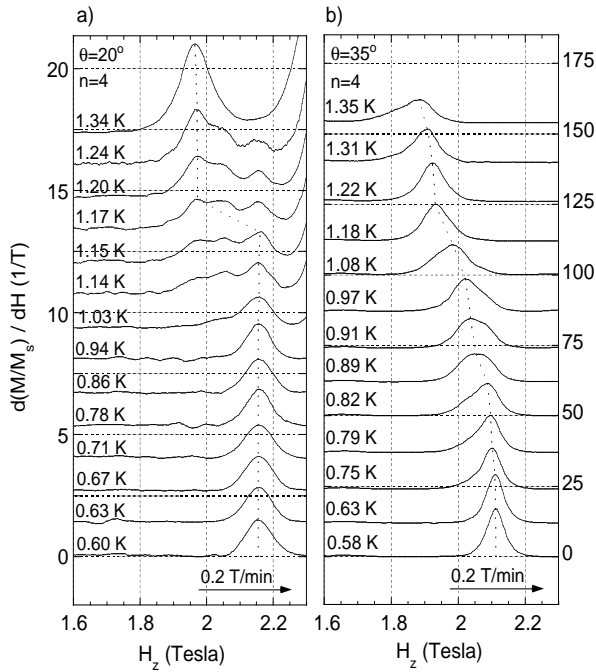}
\vskip0.5cm
\caption{Field derivative of normalized magnetization vs $H_z$ at different 
temperatures for two orientations of the applied field and magnetic easy axis: 
a) $\theta=20^\circ$, showing an abrupt crossover, and b) $\theta=35^\circ$, 
showing a smooth crossover to QT. The curves are offset for clarity. 
The dashed line marks the position of the maximum in $dM/dH$. Note that the data 
on graphs a) and b) are plotted on different scales.}
\label{fig2}
\end{center}
\end{figure}

\begin{figure}
\begin{center}
\includegraphics[width=7cm,height=7cm]{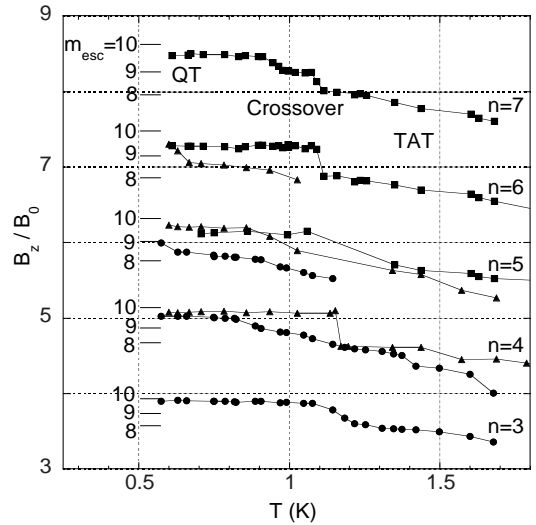}
\vskip0.5cm
\caption{Peak positions (in the units of $B_0=0.42$~T) 
vs temperature for $\theta=0^\circ$ (squares), $\theta=20^\circ$ (triangles),  
$\theta=35^\circ$ (circles). The bars on the left hand side of the graph show 
the escape levels calculated using Eq. (2). The accuracy with which the peak positions 
can be determined is approximately the size of the symbol.}
\label{fig3}
\end{center}
\end{figure}

Peak position data as a function of temperature are summarized in Fig. 3, 
which shows the values of the longitudinal field, at which the maxima of the 
peaks occur, versus temperature for the three studied orientations. 
As mentioned above, determination of the peak positions must take 
into account the internal magnetic fields in the crystal. These depend on both 
the magnetization and the crystal shape (via the demagnetization factors). 
We have used the correction coefficient $\alpha$ to determine the shift due to the 
magnetization of the sample: $B_z=H_z+4\pi \alpha M_z$. The maximum correction is 
$\Delta B_z=8\pi \alpha M_s=0.064$~T and is relatively small on the scale of the plot in Fig. 3. 
The bars on the left hand side of the figure show the 
escape levels calculated by using Eq. (2), with parameters from spectroscopic data 
\cite{EPR3,NS}. 
The correspondence between these levels and the observed peak positions is
remarkably good, given the approximations involved in the analysis.

By analyzing this graph, we can make following observations. First, for larger angles, 
and therefore higher transverse field, peaks with lower indices can be observed in 
the experimental time window.  The lowest step observed for $\theta=0^\circ$ 
is $n=5$, for $\theta=20^\circ$ it is $n=4$, for $\theta=35^\circ$ it is $n=3$. 
This is consistent with the idea 
that the transverse field promotes tunneling and lowers the effective anisotropy 
barrier. We find that there is greater amplitude in lower lying peaks as the 
transverse field is increased. Second, two regimes can be distinguished: the 
high temperature regime, where the peaks gradually shift to higher fields with 
decreasing temperature, and the low temperature regime, where the peak positions 
are constant. We associate the first regime with the TAT and the second with pure QT. 
Third, the form of the crossover between these two regimes depends on the 
applied field. For each sample orientation, peaks with lower indices (smaller $H_z$) 
show a more abrupt crossover between TAT and QT than peaks with higher indices 
(compare peaks $n=6$ and $n=7$ for $\theta=0^\circ$, or $n=4$ and $5$ for 
$\theta=20^\circ$, 
or $n=3$ and $4$ for $\theta=35^\circ$). 

\vspace{1cm}
\begin{figure}\begin{center}
\includegraphics[width=7cm]{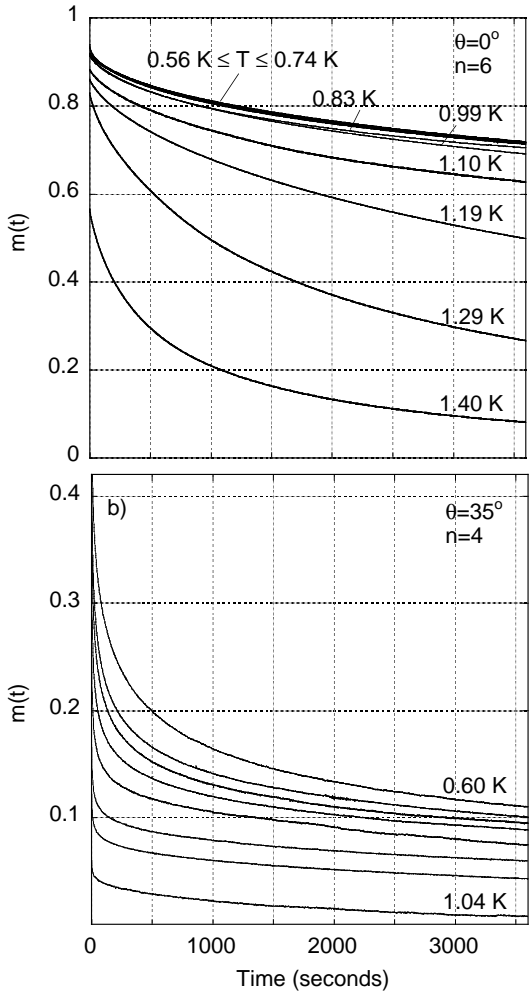}
\vskip0.5cm
\caption{Relaxation of the magnetization vs time at different temperatures 
for a) $n=6$, $\theta=0^\circ$, showing a crossover to a 
quantum regime at approximately $1$ K, and b) $n=4$, $\theta=35^\circ$, 
showing no temperature independent regime. $m(t)$ is a reduced magnetization: 
$m(t)=(M_s-M(t))/2M_s$. In a) the five curves below $0.74$~K overlap 
($0.56$~K, $0.58$~K, $0.63$~K, $0.68$~K, $0.74$~K). These curves can be
fit with $m(t)=m_0\exp((-t/\tau)^\beta)$, where 
$m_0=0.94\pm 0.01$, $\tau=(5.45\pm 0.15)\cdot 10^4$ s, $\beta=0.48\pm 0.02$. 
The fit overlaps the data. In b) the unmarked curves 
from top to bottom correspond to $T=0.68$~K, $0.70$~K, $0.75$~K, $0.83$~K, $0.91$~K, 
$0.95$~K.}
\label{fig4}
\end{center}
\end{figure}

The crossover from TAT to QT is also evident in magnetization relaxation 
measurements. In these experiments the sample was first saturated ($M=-M_s$), 
then the field was ramped (at $0.2$~T/min) to a certain value and held constant 
for 1 h, during which the magnetization was measured as a function 
of time. Figure 4 shows two sets of relaxation curves measured at $0^\circ$ and 
$35^\circ$ 
at the fields where peaks $n=6$ and $n=4$, respectively, occur at the 
lowest temperature. 
For $n=6$, $\theta=0^\circ$ below approximately $1.1$~K, the 
relaxation 
curves are spaced very closely, i.e., the relaxation rate almost does not change, 
while at higher temperature it changes significantly. 
This temperature corresponds to the crossover temperature seen in Fig. 3 -- consistent
with pure QT. 
In contrast, for the peak $n=4$, $\theta=35^\circ$, the magnetization relaxation rate changes 
significantly as the temperature decreases in the entire studied range. Relaxation curves 
can be fit by a stretched exponential function $m(t)=m_0\exp(-(t/\tau)^\beta)$,
where $\beta \approx 0.4 - 0.6$.  This form of relaxation has been observed 
previously in Fe$_8$ \cite{Ohm} and in Mn$_{12}$ \cite{Barbara}, although it is not completely 
understood \cite{Nonexp}.

In summary, we have presented new low temperature magnetic studies of thermally assisted and pure 
quantum tunneling in Mn$_{12}$. The crossover between these two regimes was found to be either 
abrupt or gradual, depending on the magnitude and orientation of applied magnetic field. Higher 
longitudinal and transverse fields broaden the crossover, consistent with a recent model 
\cite{Garanin}. 
We have also shown that below the crossover temperature the magnetization relaxation becomes 
temperature independent. 
We note that the measured crossover temperature ($\sim 1.1$ K) is significantly higher than 
predicted ($0.6$ K) \cite{Polyhedron}. This may be due to an intrinsic mechanism promoting 
tunneling in Mn$_{12}$ such as a transverse anisotropy.
Further studies of this crossover will lead to a better understanding of the 
mechanisms of relaxation in Mn$_{12}$.
 
This work was supported by NSF-INT (Grant No. 9513143) and NYU.

\end{multicols}

\begin{references}
\bibitem{Chudnovsky}
see, E. M. Chudnovsky and J. Tejada, {\it Macroscopic tunneling of the magnetic moment}  
(Cambridge University Press, Cambridge, UK 1997), Chapter 7.

\bibitem{Friedman}
J. R. Friedman, M. P. Sarachik, J. Tejada, and R. Ziolo, Phys. Rev. Lett. {\bf 76}, 
3830 (1996); L. Thomas, F. Lionti, R. Ballou, D. Gatteschi, R. Sessoli, and B. Barbara, 
Nature {\bf 383}, 145 (1996).

\bibitem{Ohm}
C. Sangregorio, T. Ohm, C. Paulsen, R. Sessoli, D. Gatteschi, Phys. Rev. Lett. 
{\bf 78}, 4645 (1997).

\bibitem{Sessoli}
R. Sessoli, D. Gatteschi, A. Caneschi, and M. A. Novak, Nature {\bf 365}, 141 (1993). 

\bibitem{Novak}
M. Novak and R. Sessoli, in {\it Quantum Tunneling of Magnetization-QTM'94}, 
ed. by L. Gunther and B. Barbara (Kluwer Publishing, Dordrecht, 1995) p. 171; 
B. Barbara {\it et al.}, JMMM {\bf 140-144}, 1825 (1995).

\bibitem{Hernandez}
J. M. Hernandez {\it et al.}, Europhys. Lett. {\bf 35}, 301 (1996).

\bibitem{Wernsdorfer}
W. Wernsdorfer and R. Sessoli, Science {\bf 284}, 133 (1999).

\bibitem{EPR1}
S. Hill {\it et al.}, Phys. Rev. Lett. {\bf 80}, 2453 (1998). 

\bibitem{EPR2}
M. Hennion {\it et al.}, Phys. Rev. B {\bf 56}, 8819 (1997).

\bibitem{EPR3}
A. L. Barra, D. Gatteschi, and R. Sessoli, Phys. Rev. B {\bf 56}, 8192 (1997).

\bibitem{NS}
I. Mirebeau {\it et al.}, Phys. Rev. Lett. {\bf 83}, 628 (1999).

\bibitem{CCNS}
Y. Zhong {\it et al.}, J. Appl. Phys. {\bf 85}, 5636 (1999).

\bibitem{Crossover}
E. M. Chudnovsky and D. A. Garanin, Phys. Rev. Lett. {\bf 79}, 4469 (1997).

\bibitem{Crosstheory}
G.-H. Kim, Phys. Rev. B {\bf 59}, 11847 (1999); G.-H. Kim, Europhys. Lett. {\bf 51}, 
216 (2000); H. J. W. M\"{u}ller-Kirsten, D. K. Park, and J. M. S. Rana, Phys. Rev. B 
{\bf 60}, 6662 (1999), and references therein.

\bibitem{Note}
The analogy to phase transitions is a purely formal one as discussed in 
\cite{Crossover}. 
The first-order, second-order terminology for the escape rate crossover is originally 
due to Larkin and Ovchinnikov \cite{Larkin}.

\bibitem{Larkin}
A. I. Larkin and Y. N. Ovchinnikov, Sov. Phys. JETP {\bf 59}, 420 (1984).

\bibitem{OurEPL}
A. D. Kent {\it et al.}, Europhys. Lett. {\bf 49}, 512 (2000).

\bibitem{WernsdorferEPL}
W. Wernsdorfer {\it et al.}, Europhys. Lett. {\bf 50}, 552 (2000).

\bibitem{HallBar}
A. D. Kent, S. von Molnar, S. Gider, and D. D. Awschalom, J. Appl. Phys. {\bf 76}, 
6656 (1994).

\bibitem{Lis}
T. Lis, Acta Cryst. B {\bf 36}, 2042 (1980).

\bibitem{Barbara}
L. Tomas, A. Caneschi, and B. Barbara, Phys. Rev. Lett. {\bf 83}, 2398 (1999).

\bibitem{Nonexp}
N. V. Prokof'ev and P. C. E. Stamp, Phys. Rev. Lett. {\bf 80}, 5794
(1998); E. M. Chudnovsky, Phys. Rev. Lett. {\bf 84}, 5676 (2000); N. V Prokof'ev and P. C. E. 
Stamp, {\it ibid.}, {\bf 84}, 5677 (2000); W. Wernsdorfer, C. Paulsen, and R. Sessoli,
{\it ibid.}, {\bf 84}, 5678 (2000).

\bibitem{Garanin}
D. A. Garanin, X. Hidalgo, and E. M. Chudnovsky, Phys. Rev. B {\bf 57}, 13639 (1998).

\bibitem{Polyhedron}
L. Bokacheva, A. D. Kent, and M. A. Walters, Polyhedron (to be published).
\end{references}
\end{document}